\documentclass[12pt]{article}

\usepackage{amsmath}
\usepackage{amsfonts}
\usepackage{amssymb}
\usepackage{graphicx}

\begin{document}

\newtheorem{problem}{Problem}
\newtheorem{definition}{Definition}
\newtheorem{lemma}{Lemma}
\newtheorem{proposition}{Proposition}
\newtheorem{corollary}{Corollary}
\newtheorem{example}{Example}
\newtheorem{conjecture}{Conjecture}
\newtheorem{algorithm}{Algorithm}
\newtheorem{theorem}{Theorem}
\newtheorem{condition}{Condition}
\newtheorem{exercise}{Exercise}

\newcommand{\la}{{\lambda}}
\newcommand{\eps}{{\varepsilon}}
\newcommand{\bfA}{{\mathbf{A}}}

\newcommand{\ls}[1]
   {\dimen0=\fontdimen6\the\font \lineskip=#1\dimen0
\advance\lineskip.5\fontdimen5\the\font \advance\lineskip-\dimen0
\lineskiplimit=.9\lineskip \baselineskip=\lineskip
\advance\baselineskip\dimen0 \normallineskip\lineskip
\normallineskiplimit\lineskiplimit \normalbaselineskip\baselineskip
\ignorespaces }

\title{On the optimality of universal classifiers for  finite-length
 individual test sequences.}

\author{Jacob   Ziv\\
                    Department of Electrical Engineering\\
                     Technion--Israel Institute of Technology\\
Haifa 32000,  Israel}

\date{\it November 5, 2011}

\maketitle

\ls{1.5}

\begin{abstract}

 We consider pairs of finite-length individual sequences that are realizations of unknown, finite alphabet, stationary sources in a class
 $ {M}$ of sources with vanishing memory (e.g. stationary Markov sources).

The task of a universal classifier is to decide whether the two sequences are emerging from the same source or  are emerging from two distinct sources in $ {M}$, and it has to carry this task without any prior knowledge of the two underlying probability measures.

Given a fidelity function and a fidelity criterion, the probability of classification error for a given universal classifier is defined.

Two universal classifiers are defined for pairs of $N$ -sequence: A  "classical" fixed-length (FL) universal classifier and an alternative variable-length (VL) universal classifier.

 Following Wyner and Ziv (1996)  it is demonstrated that  if the length of the individual sequences $N$ is
  such that ${N}< {N_0}^{1-\epsilon}$, any universal classifier will fail  with high probability .
 ( $N_0$ is a constant that is being determined by  the parameters of  $M$ and $\epsilon$ is an arbitrarily small positive number).

It is demonstrated that for $N> {N_{0}}^{1+\epsilon}$, the classification error relative to either one of the two classifiers tends to zero as the length of the sequences tends to infinity
 However, the probability of classification error that is associated with the VL-universal classifier is uniformly smaller or equal to the one that is associated with  the "classical" fixed length universal classifier, for any finite length.

${\bold {Index ~terms:}}$ universal classification,
 universal data-compression.
\end{abstract}
\ls{1.5}
\section{Introduction, notations and definitions}
 A device called a ${\bf classifier}$  (or discriminator)  observes two
$N$-sequences whose probability laws are  $Q$ and $P$
respectively ( $Q$ and $P$ are defined on doubly infinite sequences in
a finite alphabet $ \bfA $). Both $Q$ and $P$ are unknown. The
classifier's task is to decide whether $P=Q$, or $P$ and $Q$ are
sufficiently different according to some appropriate criterion
$\Delta$. If the classifier has available an infinite amount of
training data (i.e. if $N$ is large enough), this is a simple
matter. However, here we study the case where $N$ is finite.

Following [2], consider random sequences from a finite alphabet $\bfA$, where
$|{\bfA}|=A <\infty $. Denote $\ell$ vectors from $\bfA$ by
$z^{\ell}=z_1,...z_{\ell} \in {\bfA}^{\ell}$, and use upper case
$Z$'s to denote random variables. When the superscript is clear from
the context, it will be omitted. Similarly, a substring
$Z_{i},\dots, Z_{j}; -\infty \leq i<j \leq + \infty$ is denoted by
$Z_{i}^{j}$.

Let a class of "vanishing memory" processes $M$ be defined as
follows:
Given a positive constants  $0<\delta<1$
and a positive integer $\ell$, let
 $M=M_{k_0,\alpha,\beta,\delta,\ell,R}$ be the set of probability
measures on doubly infinite sequences from the set $\bfA$, with the
following properties:
\begin{itemize}
\item[A)]
Positive transitions property:
$$
P (X_1 = z_1 | X^0_{-\infty} = z_{-\infty}^0,  X_2^{\infty}
=z_2^{\infty}) \ge \delta > 0
$$
for all sequences of $z_{-\infty}^{\infty}$, for every $P \in M$, where $ 0< \delta<1$.

\item[B)]
Strong Mixing condition (following [2], Eq.~(9)):\\
Let $\{X_i\}, - \infty < i < \infty$, be  a random sequence with
probability law $P \in M$. We further assume that $\{X_i\}$ is a
stationary ergodic process where every member in $M$
satisfies the following condition:

\begin{condition} Let $\sigma
(X_i^j;-\infty \le i,j \le + \infty)$ be the $\sigma$-field
generated by the subsequence $X_i^j$.  Then, there exist  integers $\beta >1$ and
$k_o$, such that for all $k \ge k_0$, all $A\in \sigma (
X_{-\infty}^0)$ and all $B\in \sigma (X_k^\infty)$
\begin{equation}
\label{eq:2} \frac{1}{\beta} \le \frac{P(B)}{P(B|A)} \le \beta
\end{equation}
for $P(A), P(B) > 0$.
\end{condition}
\item[C)]
Let $\alpha<1$ be an arbitrarily small positive number.
For every ${ P} \in M$,

$P[ X_1^N: P[X_{1}^{n}\leq 2^{-{n}R}]\leq \alpha$ for any integer $n\geq \ell$

Also, there exists a large enough integer $n_0$ such that for any positive integer $\ell \geq n_0$,

$P[ X_1^N: P[X_{1}^{\ell}\leq 2^{-{n}R}]\leq \frac{1}{\ell}$.

\end{itemize}
The constants $k_0,\alpha, \beta$, $R$,$n_0$ and $\ell$ do not depend on $P$.

The condition in B) is reminiscent of $\phi$-mixing but is not
identical to it. We remark that if $P$ is any irreducible, aperiodic
finite-order Markov process, this condition will be satisfied.
Furthermore, the ``positive transitions" condition may be guaranteed
by dithering prior to the classification process, without violating
the strong mixing condition. The condition in C) is satisfied by any
ergodic process for some $\ell$ and $n_0$, by the Asymptotic Equipartition
Property (AEP) of information theory.

\section{ The normalized KL Fidelity Function}

Let the normalized $N$-th order K-L divergence between $Q$ and $P\in
M$ be:
$$
D_{KL,n} (Q\| P)=\frac{1}{n}KL(Q^{n}\|P^{n}) \triangleq \frac{1}{n}
\sum_{{\bf Z}\in {\bf A}^n} Q({\bf Z}) \log \frac{Q({\bf Z})}{P({\bf
Z})}
$$
where $Q^{n},P^{n}$ are the $n$-th  dimensional marginal measures of
$Q,P$, and $KL(*\|*)$ denotes the conventional Kullback-Leibler
divergence.
 Logarithms are taken  on base~2 and obey $0 \log 0\equiv 0$.

Finally, let
$$D_{KL} (Q\| P)=\liminf_{n \to \infty}D_{KL,n} (Q\| P)$$

Formally, given an $N$-sequence ${\bf Y}$ which is a realization of
$Q$ and another $N$-sequence ${\bf X}$ which is a realization of
$P$, we define a classifier $f_c$ (c-for ``classifier") as  a
mapping of $({\bf X},{\bf Y})$ to $\{0,1\}$,
$$f_c: {\bf A}^{2N}\times M \to \{0,1\}$$
 where $f_c=1$ declares $Q$ to be different from
$P$, $f_c=0$ means $Q=P$.

 The probability of classification error that is associated with any finite class $\hat M \subset M$   is defined as follows:

Let $X$ be a realization of $P_i\in \hat M$ and $Y$ be a realization
of some $P_j \in \hat M$. Then,

${ \la}_{N,n}(P_i,\hat M,f_c,\Delta) = P_r[ (\bold X,\bold Y)
: \text{\ either\ }f_c({\bf X},{\bf Y}) = 1~ and~ P_j=P_i$, or
\begin{equation}
~for ~some ~P_j: D_{KL,n}(P_j\|P_i)\ge \Delta, f_c ({\bf X}_i, {\bf
Y}_j) \equiv 0
\end{equation}
where $\Delta$ is a fidelity criterion.

Also, let
\begin{equation}
\la_{N,n}(\hat M, f_c,\Delta)=
 \sup_{P_i\in  \hat M}\la_{N,n}(P_i,\hat M, f_c,\Delta)
\end{equation}
Finally,
\begin{equation}
\la(\hat M,f_c,\Delta)=\limsup_{n\to \infty}\limsup_{N\to \infty}{\la_{N,n}(\hat M,f_c,\Delta)}
\end{equation}
A universal classifier that achieves  $\la(\hat M,f_c,\Delta)=0$, for every $\hat M \subseteqq M$  and any arbitrarily small positive $D_{KL} (Q\| P)\geq \Delta>0$ is said to be   ${\it asymptoticall}$ optimal
over the class $M$.

 It is demonstrated that a "classical" fixed length (FL) universal classifier that, given $\bold X$ and $\bold Y$,
  generates $\ell-th $ order empirical approximations of $P_i$ and $P_j$ and based on that, an empirical approximation of $D_{KL,n}(P_j\|P_i)$, ia indeed asymptotically optimal.

However, some  asymptotically optimal classifiers may be better than others for non-asymptotic length values $N$ in the sense that they yield a smaller classification error than others.

 It follows from the definition of the class $M$ below that for any probability measure $P\in M$, the probability that any  substring  of length $n\geq \ell$ or more for which

$ P(X_1^n)< \frac {1}{N_0}; N_0=2^{R\ell}$  appears in a sequence $X_1^N$ which is a realization of $P$,
 is  small (no larger than $\alpha$).

Following [2] it is first shown that $ \la_{N,n}(\hat M, f_c,\Delta)\approx 1$ for any
classification function $f_c({\bf X},{\bf Y})$ over some classes $\hat M$ as long as the length of the test sequences

$\bf X$ and $\bf Y$ satisfy:$N \leq 2^{(R-\epsilon)\ell}$ where $\epsilon$ is an arbitrarily small positive number.

Consider now the set $S_{L_{max},\ell}(\bold X,\bold Y)$ of all strings $Z_1^i \in {\bold A}^i$ that for some positive integer $i:\ell \leq i\leq L_{max}$ appear in both $\bold X$ and $\bold Y$.

 In the case of a FL universal classifier we  limit our attention to cases where $f_c(\bold X,\bold Y)=f_c(S_{L_{max}=\ell,\ell}(\bold X,\bold Y))\triangleq f_{c,FL}(\ell)$.

 It is demonstrated that if $N \geq 2^{(R+\epsilon)\ell}$  an FL classifier $f_{c.FL}(\ell)$  yields a small classification error $ \la_{N,\ell}(\hat M, f_{c,FL}(\ell),\Delta)$ if $\inf_{P_j,P_i\neq P_j \in \hat M}D_{KL,\ell}(P_j\|P_i)\ge \Delta$, the classification error vanishes as $\ell$is increasing and therfore the proposed FL classifier is asymptotically optimal.

It is next demonstrated that a variable-length (VL) asymptotically optimal universal classifier,
$f_c(S_{L_{max},\ell}(\bold X,\bold Y))\triangleq f_{c,VL}(L_{max})$,
may
outperform the classical FL classifier, for non asymptotic values of $N$.

The proposed classifier is reminiscence of [1] and utilizes the whole set $S_{L_{max},\ell}(\bold X,\bold Y)$ rather than the set of its prefixes of length $\ell$, $S_{L_{max}=\ell,\ell}(\bold X,\bold Y) $ as above.
\newpage

Define,

$$D_{VL,\ell,L_{max}} (P_j\| P_i) \triangleq
\sum_{Z_1^{\ell}\in {\bf A}^{\ell}} P_j(Z_1^{\ell})\max_{\ell\leq k\leq L_{max}:P_j(Z_1^k) \geq \frac{1}{N_0}}\frac{1}{\ell}\log \frac{P_j({Z_1^k })}{P_i(Z_1^k)}$$

Clearly, $D_{VL,\ell,L_{max}} (P_j\| P_i)\geq  D_{KL,\ell} (P_j\| P_i)$  if $P_j \neq P_i$, $D_{VL,\ell, L_{max}} (Q\| P)=  D_{KL,\ell} (P_j\| P_i)\equiv 0$  if $P_j=P_i$, and by the AEP property , $\liminf_{\ell \to \infty}D_{VL,\ell, L_{max}} (P_j\| P_i)=D_{KL} (P_j\| P_i)$

It is demonstrated that if $N \geq 2^{(R+\epsilon)\ell}$,  a VL classifier $f_{c.VL}(L_{max})$  yields a small classification error $ \la_{N,\ell}(\hat M, f_{c,VL}(L_{max}),\Delta)$ if $\sup_{P_i,P_i\neq P_j \in \hat M}D_{VL,\ell,L_{max}}(P_j\|P_i)\ge \Delta$.

Thus, $ \la_{N,\ell}(\hat M, f_{c,FL}(\ell),\Delta)$ dominates
$\la_{N,\ell}(\hat M,f_{c,VL}(L_{max}),\Delta)$ .

We give an example where $\la_{N,\ell}(\hat M,f_{c,VL}(L_{max}),\Delta)$
is small for $N\geq 2^({\ell}R+\epsilon)$ although $D_{KL,\ell} (P_j\| P_i)\leq \Delta$, thus yielding
$ \la_{N,\ell}(\hat M, f_{c,FL}(\ell),\Delta)\approx 1$.

\section{Statement of results}

In the following converse theorem it is demonstrated that
 no efficient classification is
possible.

Following the proof of Theorem 6 in [2, Theorem 6], we get the following
converse theorem:

\begin{theorem}:
\label{the-1} Let $ \hat M \subset M$
. Then, for all $ \epsilon, \Delta > 0$ and all
$R\in [0, \log A]$,
there exists a $\delta_0 = \delta_0 (\alpha, \eps, \Delta, R)$
(sufficiently small) and an ${\ell}_0 $ sufficiently large
 such that for all $\ell \geq {\ell}_0$ any discriminator $f_c$
on $ \hat M \subset M$ with parameters $N,\Delta$ for which $N \le 2^{(R-\epsilon)\ell_0}$, must satisfy  $ \la_{N,\ell_0}(\hat M, f_c,\Delta)=1$
\end{theorem}
\newpage
\paragraph{Proof of Theorem~1:}

$${\bf Construction ~of~ \hat M} ~(following ~[2,Appendix II)$$
By Lemma A1 in [2], for any positive number $0<R< \log A$, there exists a collection of cyclic subsets
${ A}_i$ of $\ell$-vectors from $[0,1]^{\ell}$,
each of size $2^{{R}\ell}$, and where, for some $\beta_0(0<\beta_0<1/2)$ the Hamming distance between any $x\in A_i,y\in A_j;(i\neq j)$, $d_H(x,y)\geq \ell\beta_0$

At time zero, choose an $\ell$-vector from ${\bold A}^{\ell}$ with a uniform distribution
on a cyclic set ${ A}_i$ . Repeat this $\ell$-vector $\nu$ times to create a ${\nu}\ell$ vector, where $\nu\geq \ell$ is a positive integer.

Next, add a ${\nu}'$-vector  consisting of the first ${\nu}'$ elements in the first vector chosen. Say that
${\nu}'$ is uniformly distributed on $[1,\ell]$. Since the sets $A_i$ are cyclic, any length $\ell$ substring of this vector belong to $A_i$. Thus,we have defined a random $({\nu}\ell+{\nu}')$ vector. The process
$ P$ is the concatenation of these sequences with a random-phase uniformly distributed between $0$ and
$({\ell}{\nu}-1)$, and dithered by the additional modulo 2 of an i.i.d. "noise" vector
$\bold W$ with $P_r(W_{i}=1)=\delta, P_r(W_{i}=0)=1-\delta$  [2, page 346].

It follows that for any $P_j\neq P_i; P_i, P_j \in \hat M$,
\begin{equation}
\log \frac{1-\delta}{\delta} > D_{KL,\ell}(P_j\|P_i)>\frac {\nu -1}{\nu +1}\log \frac{1-\delta}{\delta}
\end{equation}
 By Lemma A1 in [2] it follows that by choosing
$\delta$ to be small enough, and for any $P_i,P_j \in  \hat M ;i \neq j$ the divergence $D_{KL,\ell}(P_j\|P_i)
$, can be made arbitrarily large.
At the same time, the number of processes in $\hat M_{\ell}$ is at least $2^{2^{(R-\epsilon)\ell}}$ while there are only $2^N$  $X$ sequences to cope with $\hat M$, and by derivation similar to those of  [2],
leading to to the conclusion that $\la_{N,\ell}(\hat M, f_{c}=1$
 if $N < 2^{(R-\epsilon)\ell}$, for any classifier $f_c$, even if the measure $P_j$ that governs ${\bf Y}$,  is given.

We state now some mathematical preliminaries that are derived directly
from  [2,Eq.67, Lemma 5 and Lemma 6].

Let  $\ell$ be a positive integer, let $Z\in {\bold A}^{\ell}$  and let ${\bold X}^*=X_1^{\infty}$

Also, let

$ N(Z|{\bold X}^*)\equiv$ smallest $k$ such that ${X}_{k}^{k+\ell}(j)=Z;1\leq k$
\
\begin {lemma}
For~ an~ arbitrarily~small~ ${\epsilon}_0<0$ ~and~ any~ distribution $P \subset M$
\end {lemma}

$P[{\bold X}^*:\frac{1}{\ell}\log \hat N(Z|{\bold X}^*)\geq \frac{1}{\ell}\log \frac{1}{P(Z)}+\epsilon_0]
\leq \beta (\nu_{0}\gamma_{0})^{\ell}+2^{-\epsilon_{0}{\ell}}$

where $\gamma_0 =1-(A-1)\delta$ and $\nu_0>1$ satisfies ${\gamma_0}{\nu_0}<1$.

Also, for $Z\in {\bold A}^{\ell}$,
 let ${\bold X'}^*$  consist of $K$ independent
 drawings of ${\bold X}^*$-vectors  with distribution $P\in M$, ${\bold X}^*={\bold {X}^*(1)}, {\bold {X}^*(2)},...{\bold {X}^*(K)}$.

let

$\hat N(Z|{\bold X'}^*)\equiv$ smallest $k$ such that ${X}_{k}^{k+\ell-1}(j)=Z;1\leq k
~over ~all~ 1\leq j\leq K$.

\begin {lemma}
For~an~ arbitrary ${\epsilon}_0<0$ ~and~ any~ distribution $P \subset M$
\end {lemma}

$P[{\bf X'}^*:|\frac{1}{\ell}\log \hat N(Z|{\bold X'}^*)-\frac{1}{\ell}\log \frac{1}{P(Z)}|>\epsilon_0, for~all~ Z\in {\bold A}^{\ell}]$

$\leq K\beta (\nu_{0}\gamma_{0})^{\ell} +2^{-\epsilon_{0}{\ell}})
 + 2^{-\epsilon_{0}(K -\log A){\ell}}
 \leq K(2\beta +1)2^{-c_0({\epsilon_0}){\ell}}$

where $\gamma_0 =1-(A-1)\delta$ and $\nu_0>1$ satisfies ${\gamma_0}{\nu_0}<1$ and where
$c_0({\epsilon_0})=min({{\epsilon}_0},\log \frac{1}{(\nu_{0}\gamma_0)})$.

Consider now the training sequence ${\bf X}=X_1^{K(N+k_0)}$ and generate from
${\bf X}$ the sequence ${\bold X'}={ X}_1^{KN}={X}_1^{N+k_0}, {X}_{N +k_0+1}^{2N},...{X}_{j(N+k_0)+1}^{(j+1)N'},
...,{X}_{(K-1)(N_0+k_0)+1}^{(K)N_0}$.

Observe that ${\bf  X'}$ consists of $K$ $N$-vectors interlaced with guard spaces of $k_0$ letters each.
By the  vanishing memory property of the class $M$, these $K$ vectors are "almost" independent.

Then,  by Lemma 2 and by Condition C of the class $M$,
\newpage
\begin {lemma}
For~any~ arbitrarily~ small ~${\epsilon}_0 >0$,~ any~$P \subset M$~ and ~any $\ell \geq {\ell}_0$~ where
${\ell}_0$ ~is~ defined ~by ~Condition~ C~of~ the~class~$M$
\end {lemma}

$P[{\bf X'}:|\frac{1}{\ell}\log \hat N(Z|{\bold X'})-\frac{1}{\ell}\log \frac{1}{P(Z)}|>\epsilon_0,~for~ all~ Z
 \in {\bold A}^{\ell}$

$  \leq {K{\beta}^K}[2\beta +1)2^{-c_0({\epsilon_0}){\ell}}]+\frac {1}{\ell}$

 Let the empirical measures $\hat P_{\bold X}(Z_{1}^{\ell})$ and $\hat Q_{\bold Y}(Z_{1}^{\ell})$
 be based on the recurrence time of $Z_1^{\ell}$ in $\bold X'$ and in $\bold Y'$ respectively  where
\begin{equation}
 \hat Q_{\bold Y'}(Z_1^{\ell})=\frac{1}{ \hat N(Z_1^{\ell}|\bold Y')}
\end{equation}
and
\begin{equation}
 \hat P_{\bold X'}(Z_1^{\ell})=\frac{1}{ \hat N(Z_1^{\ell}|\bold X')}
\end{equation}
(Note that these empirical measures are not necessarily normalized probability measures).

Then, by Lemma 3 and for test sequences of length ${ N'} =K [2^{{(R+\epsilon}_0){\ell}}+k_0]$

\begin {lemma}
For~any~ arbitrarily~ small ~${\epsilon}_0 >0$, ~any~$P \subset M$,~and~$Q \subset M$~and ~any $\ell \geq {\ell}_0$~ where~${\ell}_0$ ~is~ defined ~by ~Condition~ C~of~ the~class~$M$
\end {lemma}

$Q[{\bf Y'}:|\frac{1}{\ell}\log \hat Q_{\bold Y}(Z_1^{n})-\frac{1}{\ell}\log Q(Z)|>\epsilon_0,~for~ all~ Z
 \in {\bold A}^{\ell}$

$ \leq {K{\beta}^K}[2\beta +1)2^{-c_0({\epsilon_0}){\ell}}]+\frac {1}{\ell}$

and,

$P[{\bf X'}:|\frac{1}{\ell}\log \hat P_{\bold X}(Z_1^{\ell})-\frac{1}{\ell}\log \frac{1}{P
(Z)}|>\epsilon_0,~for~ all~ Z
 \in {\bold A}^{\ell}$

$\leq {K{\beta}^K}[2\beta +1)2^{-c_0({\epsilon_0}){\ell}}]+\frac {1}{\ell}$
\newpage
\paragraph{{\bf A  Fixed-length (FL)Universal Classifier}}
~~~~~~~~~~~~~~~~~~~~~~~~~~~~~~~~~~~~~~~~~~~~~~~~~~~~~~~~~~~~~~~~~~~~~~~~~~~~~~~~~~~~

It is demonstrated now that if $ D_{KL,\ell} (Q\| P)>\Delta$, an FL universal classifier yields a small classification error for $N\geq N'=2^{R(\ell+{\epsilon})}$.

Let the two training sequences be $\bf Y=Y_1^{2K(N+k_0)+\ell};\bf X=X_1^{2K(N+k_0)+\ell}$ where
 $K$ is the parameter that appears in the derivation of Lemma 3 above.

Generate  ${\bf Y'}$ from $ Y_1^{K(N+k_0)}$  and $ \bf X'$ from $ X_1^{K(N+k_0)}$ as in Lemma 3 above.

 Also, let

  $\hat q_{\bold Y'}(Z_1^{\ell},N)= \hat Q_{\bold Y'}(Z_1^{\ell})~ if~\hat Q_{\bold Y'}(Z_1^{\ell})\geq 2^{-\ell(R+{\epsilon}_0)}
  ~
  ~else~ \hat q_{\bold Y'}(Z_1^{\ell},N)=(\delta)^{\ell}$

and

\begin{equation}
  \hat P_{\bold X'}(Z_1^{\ell},N)= \hat P_{\bold X'}(Z_1^{\ell})~ if~\hat P_{\bold X'}(Z_1^{\ell})\geq 2^{-\ell(R+{\epsilon}_0)}
  ~
  ~else~ \hat p_{\bold X'}(Z_1^{\ell},N)=(\delta)^{\ell}.
\end{equation}

Let

$d({\bf X,\bf Y})=\frac{1}{K(N+k_0)}\sum_{i=K(N+k_0)+1}^{{2K(N+k_0)}}d_i({\bf X, \bf Y})$

where
\begin {equation}
d_i(\bf X, \bf Y)=\log \hat p_{\bold X'}(X_{i(\ell)+1}^{(i+1)\ell})
-\log \hat  q_{\bold Y'}(X_{i(\ell)+1}^{(i+1)\ell})
\end {equation}

By Lemma 4 and by Eq.(7) and the properties of $M$, for $D_{KL,\ell}(Q\| P)\geq 2\Delta$, 			
\begin{equation}
E_{ Q \times P} {d_{i}({\bf X, \bf Y})> D_{KL,\ell} (Q\| P)} -{2\epsilon}_0-[{K{\beta}^K}[2\beta +1)2^{-c_0({\epsilon_0}){\ell}}]+\frac {1}{\ell}]log \frac{1}{\delta}
\end{equation}
and for $Q\equiv P $,
\begin{equation}
E_{P } d_{i}({\bf X, \bf Y})\leq {2\epsilon}_0-[{K{\beta}^K}[2\beta +1)2^{-c_0({\epsilon_0}){\ell}}]+\frac {1}{\ell}]log \frac{1}{\delta}
\end{equation}
Let$f_{c,\ell}(\bf X,\bf Y)=0$ if $d(\bf X, \bf Y)\leq \Delta$

else, $f_{c,\ell}(\bf X,\bf Y)=1$

Note that by the vanishing memory property of $M$ for any $P\in M$ and any integer $j\geq k_0$,

$P[d_{i}({\bf X, \bf Y});d_{i+j}({\bf X, \bf Y})]\leq \beta P[d_{i}({\bf X, \bf Y})]P[d_{i+j}({\bf X, \bf Y})]$

and

$P[d_{i}({\bf X, \bf Y});d_{i+j}({\bf X, \bf Y})]\geq \frac {1}{\beta} P[d_{i}({\bf X, \bf Y})]P[d_{i+j}({\bf X, \bf Y})]$

Thus, by Lemma 4, by the Chernoff bound, and by Eqs. (2),(9) and (10) and $N' \geq  2K(N+k_0)+\ell$
where $N'$ is the length of both $\bf X'$ and $\bf Y'$
\begin {equation}
\la_{N',\ell}(\hat M, f_{c,FL}(\ell),{\Delta}')\leq 2^{-\frac {N}{k_0}[m(\epsilon)]}
\end{equation}
 for
 ${\Delta'}= \Delta +2[{2\epsilon}_0+[{K{\beta}^K}[2\beta +1)2^{-c_0({\epsilon_0}){\ell}}]+\frac {1}{\ell}]log \frac{1}{\delta}+\log \beta]$ and $\Delta\leq D_{KL,\ell}(Q\|P)-\epsilon$

for some $m(\epsilon)>0$.

Observe that ${\Delta'}$ approached $\Delta$ for large enough $\ell$ and $\beta$ close enough to 1.

It follows from Eq.(5) above that if $\Delta> 2\log \frac{1-\delta}{\delta}$,
$\la_{N',\ell}(\hat M, f_{c,FL}(\ell),{\Delta}')$ can  not vanish as  $N$ tends to infinity, over the  class $ M$.

On the other hand, it will be now demonstrated that a proposed  asymptotically optimal variable-length (VL) universal classifier yields a  small classification error for $N>2^{\ell (R+\epsilon)}$ and vanishes as $N$ tends to infinity, over the particular class $\hat M$ that was described in the proof of Theorem 1 above, if $2\log \frac{1-\delta}{\delta}< \Delta <4\frac {\nu -2}{\nu}\log \frac{1-\delta}{\delta}$, thus demonstrating the superiority of the (VL)-classifier
over the "classical (FL)-classifier.
\newpage
\paragraph{{\bf A  Variable-length (FL)Universal Classifier}}
~~~~~~~~~~~~~~~~~~~~~~~~~~~~~~~~~~~~~~~~~~~~~~~~~~~~~~~~~~~~~~~~~~

 For some positive integer $L_{max}>\ell$, let the two training sequences be $\bf Y=Y_1^{2K(N+k_0)+L_{max}};\bf X=X_1^{2K(N+k_0)+L_{max}}$ where
 $K$ is the parameter that appears in the derivation of Lemma 3 above.

Generate  ${\bf Y'}$ from $ Y_1^{K(N+k_0)}$  and $ \bf X'$ from $ X_1^{K(N+k_0)}$ as in Lemma 3 above.

Let

$d_{VL}({\bf X,\bf Y})=\frac{1}{K(N+k_0)}\sum_{i=K(N+k_0)+1}^{{2K(N+k_0)}}d_{i,VL}({\bf X, \bf Y})$

where
\begin {equation}
d_{i,VL}(\bf X, \bf Y)=\log \max_{j=\ell}^{\L_{max}}\hat p_{\bold X'}(X_{ij+1}^{(i+1)j})
-\log \hat  q_{\bold Y'}(X_{ij+1}^{(i+1)j})
\end {equation}

By Lemma 3,
for any distribution $P\in M$ and large enough
 $N'$ (and the associated $N$ and $L_{max}$)

$P[{\bf {X}'}:|\log N+\log { P(Z_1^{L_{i,N,K}(\bold X)})}|\leq{\epsilon_0}L_{max}
, ~for ~all~Z_1^{i}\in {\bold A}^{i};~all \ell\leq i\leq L_{max}]$

$\leq \sum_{i=1}^{L_{max}}P[{\bf {X}'}:|\log N+\log { P(Z_1^i)}|\leq{\epsilon_0}L_{max}
, ~for ~all~Z_1^{i}\in {\bold A}^{i};~all ~\ell\leq i\leq L_{max}]$
\begin{equation}
 \leq 2({L_{max}}^2) {K{\beta}^K}[2\beta +1)2^{-c_0({\epsilon_0})L_{max}}+\frac {1}{\ell}
\end{equation}
Similarly, for any distribution $Q\in M$ and large enough
 $ N$ (and the associated  $L_{max}$)

$Q[{\bf {Y}'}:|\log N+\log { Q(Z_1^{L_{i,N,K}(\bold  X|\bold Y)}}|\geq {\epsilon_0}L_{max}
, ~for ~all~~Z_1^{i}\in {\bold A}^{i};~all ~\ell\leq i\leq L_{max}]$
\begin{equation}
 \leq 2({L_{max}}^2) [{K{\beta}^K}[2\beta +1)2^{-c_0({\epsilon_0})L_{max}}]+\frac {1}{\ell}
\end{equation}
By Eqs(13), (14) and (15) and by the properties of $M$, for $D_{VL,L_{max}}(Q\| P)\geq 2\Delta$	
	
\begin{equation}
E_{P\times Q } {d_{i,VL}({\bf X, \bf Y})> D_{VL,\ell}(Q\| P)}-[2({L_{max}}^2) [{K{\beta}^K}[2\beta +1)2^{-c_0({\epsilon_0})L_{max}}]+\frac {1}{\ell}]{\log \frac{1}{\delta}}
\end{equation}
and for $Q\equiv P $,
\begin{equation}
E_{P } d_{i,VL}({\bf X, \bf Y})< [2({L_{max}}^2) [{K{\beta}^K}[2\beta +1)2^{-c_0({\epsilon_0})L_{max}}]+\frac {1}{\ell}]{\log \frac{1}{\delta}}
\end{equation}
Let $f_{c,VL}(\bf X,\bf Y)=0$ if $d(\bf X, \bf Y)\leq \Delta$

else, $f_{c,VL}(\bf X,\bf Y)=1$

Note that by the vanishing memory property of $M$ for any $P\in M$ and any integer $j\geq k_0$,

$P[d_{i,VL}({\bf X, \bf Y});d_{i+j, VL}({\bf X, \bf Y})]\leq \beta P[d_{i,VL}({\bf X, \bf Y})]P[d_{i+j,VL}({\bf X, \bf Y})]$

and

$P[d_{i.VL}({\bf X, \bf Y});d_{i+j,VL}({\bf X, \bf Y})]\geq \frac {1}{\beta} P[d_{i,VL}({\bf X, \bf Y})]P[d_{i+j,VL}({\bf X, \bf Y})]$

Thus, by the Chernoff bound, and by Eqs. (2),(13), (16) and (17),

 and $N' \geq  2K(N+k_0)+\ell$
\begin {equation}
\la_{N',\ell}(\hat M, f_{c,VL}(L_{max},{\Delta}')\leq 2^{-\frac {N}{k_0}[m(\epsilon)]}
\end{equation}
for
 ${\Delta'}= \Delta +2[{2\epsilon}_0+[2({L_{max}}^2) [{K{\beta}^K}[2\beta +1)2^{-c_0({\epsilon_0})L_{max}}]+\frac {1}{\ell}]{\log \frac{1}{\delta}}]+\log{1}{\log \beta}]$ and $\Delta\leq D_{VL,\ell}(Q\|P)-\epsilon$

for some $m(\epsilon)>0$.

Here again,  ${\Delta'}$ approached $\Delta$ for large enough $\ell$ and $\beta$ close enough to 1.

It follows by construction that for the class $\hat M \subset M$ that was used for the establishment of Theorem 1 above (converse theorem) and setting $L_{max}=2\ell$,

$D_{VL,\ell}(Q\|P)\geq 2\frac {\nu -2}{\nu}\log \frac{1-\delta}{\delta}$

Therefore, for $2 \log \frac{1-\delta}{\delta}<\Delta<4\frac {\nu -2}{\nu}\log \frac{1-\delta}{\delta}$
the variable lengh (VL) universal classifier yield a vanishing classification error for large$\ell$ and s
$\beta$ close to 1, while the the "classical" fixed length (FL) universal algorithm does not, although both are asymptotically optimal.

Thus, as claimed above, some asymptotically optimal classifiers are better then others for non-asymptotic
values of the length of the test sequences. This is somewhat reminiscence of the results in [5] where
it was demonstrated that a variable length, recurrence-time based estimator of  entropy may be more
 efficient than the "classical" fixed-length estimator for non-asymptotic values of the test sequence length.

  \end{document}